\documentclass[12pt]{article}

\usepackage{axodraw}
\usepackage[dvips]{graphicx}
\def\lsim{\:\raisebox{-0.5ex}{$\stackrel{\textstyle<}{\sim}$}\:}
\def\gsim{\:\raisebox{-0.5ex}{$\stackrel{\textstyle>}{\sim}$}\:}
\def\bmaT{\left(\begin{array}{ccc}}
\def\emaT{\end{array}\right)}
\def\bma{\left( \begin{array} }
\def\ema{\end{array} \right)}

\begin{document}
{\flushright
\hspace{30mm} KEK-TH-1056\\
\hspace{30mm} hep-ph/0512077 \\
\hspace{30mm} December 2005 \\
}
\vspace{1cm}

\begin{center}
{\Large\sc {\bf Multi-photon signatures at the Fermilab Tevatron}}
\vspace*{3mm}
\vspace{1cm}

{\large {A.G. Akeroyd}$^a$, {A. Alves}$^b$, {Marco A. D\'{\i}az}$^c$, 
{O. \'Eboli}$^b$}\\
\vspace{1cm}
{\sl
a: Theory Group, KEK, 1-1 Oho, Tsukuba, Japan 305-0801\\
\vspace*{0.2cm}
b: Instituto de F\'{\i}sica, Universidade de S\~ao Paulo, 
S\~ao Paulo, Brazil\\
\vspace*{0.2cm}
c: Departamento de F\'{\i}sica, Universidad Cat\'olica de Chile,\\
Avenida Vicu\~na Mackenna 4860, Santiago, Chile 
\vspace*{0.2cm}
}
\end{center}

\vspace{2cm}

\begin{abstract}
\noindent
Fermiophobic Higgs bosons ($h_f$) exhibiting large branching ratios to
two photons can arise in models with two or more scalar doublets
and/or triplets. In such models the conventional production mechanisms
at hadron colliders, which rely on the $h_fVV$ coupling ($V=W,Z$),
may be rendered ineffective due to severe mixing angle suppression. In this
scenario, double $h_f$
production may proceed via the complementary mechanism $qq' \to H^{\pm}
h_f$
with subsequent decay $H^{\pm} \to h_fW^*$, leading to events with up to 4
photons. We perform a simulation of the detection prospects of $h_f$
in the multi-photon ($>3$) channel at the Fermilab Tevatron and show that
a sizeable region of the ($m_{H^\pm},m_{h_f}$) parameter space
can be probed during Run II.

\end{abstract}

\newpage

\section{Introduction}

Neutral Higgs bosons with very suppressed couplings to fermions --
``fermiophobic Higgs bosons'' ($h_f$) \cite{Weiler:1987an}-- may arise in
specific versions of the Two Higgs Doublet Model (2HDM)
\cite{Haber:1979jt},\cite{Gunion:1989we} or in models with Higgs triplets
\cite{Georgi:1985nv}. Such a $h_f$ would decay dominantly to two photons,
$h_f\to \gamma\gamma$, for $m_{h_f}< 95$ GeV or to two massive gauge bosons,
$h_f\to VV^{(*)}$, ($V=W^\pm,Z$) for $m_{h_f}> 95$ GeV
\cite{Stange:1994ya,Diaz:1994pk}.  The large branching ratio (BR) for 
$h_f\to \gamma\gamma$ would provide a very clear experimental signature, 
and observation of such a particle would strongly constrain the 
possible choices of the underlying Higgs sector
\cite{Stange:1994ya,Diaz:1994pk,Barger:1992ty,Pois:1993ay,
  Akeroyd:1996hg,Barroso:1999bf,Brucher:1999tx}.

Experimental searches for $h_f$ at LEP and the Fermilab Tevatron have been
negative so far.  Mass limits have been set in a benchmark model which assumes
that the coupling $h_fVV$ has the same strength as the Standard Model (SM)
Higgs coupling $VV\phi^0$, and that all fermion BRs are exactly zero.  Lower
bounds of the order $m_{h_f}\gsim 100$ GeV have been obtained by the LEP
collaborations OPAL\cite{Abbiendi:2002yc}, DELPHI\cite{Abreu:2001ib},
ALEPH\cite{Heister:2002ub}, and L3\cite{Achard:2002jh}, utilizing the channel
$e^+e^-\to h_fZ$, $h_f\to \gamma\gamma$.  At the Tevatron Run I, the limits on
$m_{h_f}$ from the D\O\ and CDF collaborations are respectively 78.5 GeV
\cite{Abbott:1998vv} and 82 GeV \cite{Affolder:2001hx} at $95\%$ C.L., using
the mechanism $qq'\to V^*\to h_fV$, $h_f\to \gamma\gamma$, with the dominant
contribution coming from $V=W^\pm$. For an integrated luminosity of 2
fb$^{-1}$, Run II will extend the coverage of $m_{h_f}$ in the benchmark model
slightly beyond that of LEP \cite{Mrenna:2000qh},\cite{Landsberg:2000ht}. In
addition, Run II will be sensitive to the region 
$110 \,{\rm GeV} < m_{h_f} < 160
\,{\rm GeV}$ and BR$(h_f\to \gamma\gamma)>4\%$ which could not be probed at
LEP.  A preliminary search in the inclusive $2\gamma +X$ channel has been
performed with $190 \,pb^{-1}$ of Run II data \cite{Melnitchouk:2005xv}.

However, the $h_fVV$ coupling in a specific model could be suppressed relative
to the $\phi^0VV$ coupling by a mixing angle, leading to a weakening of the
above mass limits.  If this suppression were quite severe ($h_fVV/\phi^0VV
<0.1$) a very light $h_f$ ($m_{h_f}<< 100$ GeV) would have eluded the searches
at LEP and the Tevatron Run I in production mechanisms which rely upon the
$h_fVV$ coupling. Therefore it is of interest to consider other production
mechanisms for $h_f$ which may allow observable rates if the $h_fVV$
coupling is suppressed. Since the couplings $h_fVV$ and $h_fVH$ (where $H$ is
another Higgs boson in the model) are complementary, two LEP collaborations,
{\em i.e.} OPAL \cite{Abbiendi:2002yc} and DELPHI \cite{Abreu:2001ib}, also
searched for fermiophobic Higgs bosons in the channel $e^+e^-\to A^0h_f$, and
ruled out the region $m_A+m_{h_f}< 160$ GeV. However, a very light $m_{h_f}<50
$ GeV is still possible if $m_A$ is sufficiently heavy.

An alternative production mechanism which also depends on the complementary
$h_fVH$ coupling is the process $qq'\to H^\pm h_f$ \cite{Akeroyd:2003bt},
\cite{Akeroyd:2003xi}. Such a mechanism is exclusive to a hadron collider, and
can offer promising rates at the Tevatron Run II provided that $H^\pm$ is not
too far above its present mass bound $m_{H^\pm}> 90$ GeV. This alternative
experimental signature depends on the decays of $H^\pm$. In fermiophobic
models the decay $H^\pm\to h_f W^{(*)}$ can have a larger BR than the
conventional decays $H^\pm\to tb,\tau\nu$
\cite{Akeroyd:1998dt},\cite{Akeroyd:1998zr}, which leads to double $h_f$
production.

In this paper we analyze the inclusive production of multi-photon
(3$\gamma$'s or 4$\gamma$'s) final states at the Tevatron RUN II
via the mechanism:
\[
  p\bar{p}\rightarrow h_f H^\pm\rightarrow h_f h_f W^\pm \rightarrow 
                              \gamma\gamma\gamma(\gamma) +X \; .
\] 
In the 2HDM the multi-photon signature arises in the parameter space
$m_{h_f}\lsim 90$ GeV, $m_{H^\pm}\lsim 200 $ GeV, and $\tan\beta > 1$.  In
this region, $BR(h_f\rightarrow \gamma\gamma) \sim 1$ and $BR(H^\pm\rightarrow
h_f W^{*\pm}) \sim 1$, leading to a $4\gamma + \hbox{leptons or jets}$
signature. The multi-photon signature has the added virtue of being
extremely clean concerning the background contamination, 
in contrast to the conventional searches for single
$h_f$ production in the channels $\gamma\gamma + V$ and $\gamma\gamma + X$.
In the present work we show that the multi-photon signal can be observed in a
large fraction of the $m_{h_f} \otimes m_{H^\pm}$ plane at the Tevatron RUN
II.  In fact, at 3$\sigma$ level of statistical significance, the RUN
II will be able to exclude Higgs masses up
to $m_{H^\pm} \lsim 240$ GeV for very light $m_{h_f}$, or $m_{h_f} \lsim 100$
GeV for $m_{H^\pm}\approx 100$ GeV.

Our work is organized as follows. In Section 2 we give a brief introduction to
fermiophobic Higgs bosons, exhibiting the main decay channels of $h_f$ and
$H^\pm$. The possible fermiophobic Higgs production mechanisms and respective
signatures are described in Section 3. We present our analyses in Section 4
and Section 5 contains our conclusions.

\section{Fermiophobic Higgs bosons}

In this section we briefly review the properties of $h_f$.  For a detailed
introduction we refer the reader to \cite{Diaz:1994pk}, \cite{Akeroyd:1996hg},
\cite{Barroso:1999bf}, \cite{Brucher:1999tx}.  Fermiophobia can arise in i)
2HDM (Model I) and ii) Higgs triplets models.  In (i) the imposition of a
discrete symmetry together with a vanishing mixing angle ensures exact
fermiophobia at tree-level.  In (ii), gauge invariance forbids any coupling of
$h_f$ to quarks while lepton couplings are strongly constrained by neutrino
oscillation data and lepton flavour violation experiments, resulting in
approximate fermiophobia at tree-level.

\subsection{2HDM (Model I)}    

If $\Phi_1$ and $\Phi_2$ are two Higgs $SU(2)$ doublets with 
hypercharge $Y=1$, the most general $SU(2)\times U(1)$ gauge invariant
scalar potential is \cite{Gunion:2005ja}:
\begin{eqnarray}
V&=&m_{11}^2\Phi_1^{\dagger}\Phi_1+m_{22}^2\Phi_2^{\dagger}\Phi_2-
\left( m_{12}^2\Phi_1^{\dagger}\Phi_2+{h.c.}\right)+
{\textstyle{1\over2}}\lambda_1\left(\Phi_1^{\dagger}\Phi_1\right)^2
\nonumber\\ &&
+{\textstyle{1\over2}}\lambda_2\left(\Phi_2^{\dagger}\Phi_2\right)^2
+\lambda_3\left(\Phi_1^{\dagger}\Phi_1\right)
\left(\Phi_2^{\dagger}\Phi_2\right)
+\lambda_4\left(\Phi_1^{\dagger}\Phi_2\right)
\left(\Phi_2^{\dagger}\Phi_1\right)
\label{potential}\\ &&
+\left\{{\textstyle{1\over2}}\lambda_5
\left(\Phi_1^{\dagger}\Phi_2\right)^2
+\left[\lambda_6\left(\Phi_1^{\dagger}\Phi_1\right)+
\lambda_7\left(\Phi_2^{\dagger}\Phi_2\right)\right]\Phi_1^{\dagger}\Phi_2
+{h.c.}\right\} \; .
\nonumber
\end{eqnarray}
If the discrete symmetry $\Phi_1\rightarrow-\Phi_1$ is imposed the 
couplings $\lambda_6=\lambda_7=0$. However, the term proportional to
$m_{12}^2$ can remain as a soft violation of the above discrete symmetry
and still ensure that Higgs-mediated tree-level flavour changing neutral 
currents are absent \cite{Gunion:1989we}.
Note that the above 2HDM potential contains one more free parameter than
those studied in Refs.~\cite{Barroso:1999bf},\cite{Brucher:1999tx}.

The potential in eq.~(\ref{potential}) breaks $SU(2)_L\times U(1)_Y$ down 
to $U(1)_{em}$ when the two Higgs doublets acquire vacuum expectation 
values
\begin{equation}
\langle \Phi_1 \rangle = \frac{1}{\sqrt{2}}
\left(\matrix{0 \cr v_1}\right)
\,,\qquad
\langle \Phi_2 \rangle = \frac{1}{\sqrt{2}}
\left(\matrix{0 \cr v_2}\right)
\end{equation}
which must satisfy the experimental constraint
$m_Z^2=\frac{1}{2}(g^2+g'^2)(v_1^2+v_2^2)\approx(91~ \mathrm{GeV})^2$.
The minimization conditions that define the vacuum expectation values
in terms of the parameters of the potential ($\lambda_6=\lambda_7=0$)
are
\begin{eqnarray}
t_1&=&m_{11}^2 v_1-m_{12}^2 v_2+\textstyle{1\over2}\lambda_1v_1^3
+\textstyle{1\over2}(\lambda_3+\lambda_4+\lambda_5)v_1v_2^2\,=\,0
\nonumber\\
t_2&=&m_{22}^2 v_2-m_{12}^2 v_1+\textstyle{1\over2}\lambda_2v_2^3
+\textstyle{1\over2}(\lambda_3+\lambda_4+\lambda_5)v_1^2v_2\,=\,0
\end{eqnarray}
from which $m_{11}^2$ and $m_{22}^2$ can be solved in favour of $m_Z^2$
and $\tan\beta\equiv v_2/v_1$.

The neutral CP-odd Higgs mass matrix is, after using the minimization 
conditions,
\begin{equation}
{\bf M}^2_A=\left(\matrix{
m_{12}^2t_{\beta}-\lambda_5v^2s_{\beta}^2 &
-m_{12}^2+\lambda_5v^2s_{\beta}c_{\beta} \cr
-m_{12}^2+\lambda_5v^2s_{\beta}c_{\beta} &
m_{12}^2/t_{\beta}-\lambda_5v^2c_{\beta}^2
}\right)
\end{equation}
and is diagonalized by a rotation in an angle $\beta$. We define
$s_\beta = \sin\beta$, $c_\beta = \cos\beta$, and $t_\beta =\tan
\beta$.~ ${\bf M}^2_A$ has a zero eigenvalue corresponding to the
neutral Goldstone boson while its second eigenvalue is the mass of the
physical CP-odd Higgs boson $A$,
\begin{equation}
m_A^2=\frac{m_{12}^2}{s_{\beta}c_{\beta}}-\lambda_5v^2
\end{equation}
with $v^2=v_1^2+v_2^2$. The charged Higgs mass matrix is given by
\begin{equation}
{\bf M}^2_{H^{\pm}}=\left(\matrix{
m_{12}^2t_{\beta}-\frac{1}{2}(\lambda_4+\lambda_5)v^2s_{\beta}^2 &
-m_{12}^2+\frac{1}{2}(\lambda_4+\lambda_5)v^2s_{\beta}c_{\beta} \cr
-m_{12}^2+\frac{1}{2}(\lambda_4+\lambda_5)v^2s_{\beta}c_{\beta} &
m_{12}^2/t_{\beta}-\frac{1}{2}(\lambda_4+\lambda_5)v^2c_{\beta}^2
}\right)
\end{equation}
which also is diagonalized by a rotation in an angle $\beta$. It has a 
zero eigenvalue corresponding to the charged Goldstone boson, and the 
charged Higgs mass is
\begin{equation}
m_{H^{\pm}}^2=m_A^2+\frac{1}{2}(\lambda_5-\lambda_4)v^2  \; .
\end{equation}
Here we see that the charged and the CP-odd Higgs masses are independent
parameters, as opposed to supersymmetry, where the mass squared difference 
is equal to $m_W^2$ at tree level.

The neutral CP-even Higgs mass matrix is given by
\begin{equation}
{\bf M}^2_{H^0}=\left(\matrix{
m_A^2s_{\beta}^2+\lambda_1v^2c_{\beta}^2 &
-m_A^2s_{\beta}c_{\beta}+(\lambda_3+\lambda_4)v^2s_{\beta}c_{\beta} \cr
-m_A^2s_{\beta}c_{\beta}+(\lambda_3+\lambda_4)v^2s_{\beta}c_{\beta} &
m_A^2c_{\beta}^2+\lambda_2v^2s_{\beta}^2
}\right)
\end{equation}
and the two eigenvalues are the masses of the neutral CP-even Higgs
bosons $h$ and $H$. It is diagonalized by an angle $\alpha$ defined by
\begin{equation}
\sin2\alpha={{\left[-m_A^2+(\lambda_3+\lambda_4)v^2\right]s_{2\beta}}
\over
{\sqrt{
\left[m_A^2c_{2\beta}-\lambda_1v^2c_{\beta}^2+\lambda_2v^2s_{\beta}^2
\right]^2+\left[m_A^2-(\lambda_3+\lambda_4)v^2\right]^2s^2_{2\beta}
}}} \; .
\end{equation}

Fermiophobia is caused by imposing the mentioned discrete symmetry 
$\Phi_1\to -\Phi_1$ which forbids $\Phi_1$ coupling to the fermions.  
This model is usually called ``Type I'' \cite{Haber:1979jt}.  However, 
fermiophobia is erased due to the mixing in the CP--even neutral Higgs 
mass matrix, which is diagonalized by the mixing angle $\alpha$, when
both CP--even eigenstates $h^0$ and $H^0$ acquire a coupling to the 
fermions.  

The fermionic couplings of the lightest CP--even Higgs $h^0$ take the form
$h^0f\overline f \sim \cos\alpha/\sin\beta$, where $f$ is any fermion.
Small values of $\cos\alpha$ would strongly suppress the fermionic 
couplings, and in the limit $\cos\alpha \to 0$ the coupling 
$h^0f\overline f$ would vanish at tree--level, giving rise to 
fermiophobia. This is achieved if
\begin{equation}
m_A^2=(\lambda_3+\lambda_4)v^2 \; .
\end{equation}
Despite this extra constraint, the parameters $m_A$, $m_{H^{\pm}}$, and
$\tan\beta$ are still independent parameters in this model.

However, at the one-loop level, Higgs boson couplings to fermions
receive contributions from loops involving vector bosons and other Higgs 
bosons,
\begin{center}
\vspace{-30pt} \hfill \\
\begin{picture}(120,70)(0,23) 

\DashLine(30,30)(60,30){3}
\Text(20,30)[]{$h_f$}
\Photon(60,30)(80,50){3}{4.5}
\ArrowLine(80,50)(110,60)
\Text(120,60)[]{$f$}
\Photon(60,30)(80,10){-3}{4.5}
\ArrowLine(110,0)(80,10)
\Text(120,0)[]{$\overline f$}
\ArrowLine(80,10)(80,50)

\end{picture}  
$\sim \, \frac{1}{16\pi^2}(gm_W)(\frac{g^2}{8})m_f
C_0(m_h^2,0,0;0,m_W^2,m_W^2)$
\vspace{30pt} \hfill \\
\end{center}
\vspace{10pt}
where we have naively estimated the contribution of the loop with the
Passarino--Veltman function $C_0$. To get an order of magnitude of the
correction we approximate $C_0\sim 1/m_h^2$, expected in the limit of large
Higgs mass, and compare this correction with the tree-level vertex in the SM
$g_{\phi^0 ff}\sim gm_f/2m_W$. We find
\begin{equation}
\frac{\Delta g_{hff}}{g_{\phi^0 ff}}\sim \frac{g^2}{64\pi^2}
\left(\frac{m_W}{m_h}\right)^2   \; .
\end{equation}
This estimation is also applicable if $m_h\lsim m_W$ replacing $m_h$ by
$m_W$. This is a very small correction. Nevertheless, we note that the 
proper renormalization of the $\phi^0 f\bar f$ vertex involves a 
counterterm that has to be taken into account.
It is conventional to define an extreme $h_f$ in which all BRs to 
fermions are set to zero. This gives rise to benchmark BRs which are used 
in the current searches to set limits on $m_{h_f}$.

\subsection{Higgs Triplet Models}

Fermiophobia (or partial fermiophobia) can arise for scalar 
fields in isospin $I=1$ triplet
representations. Gauge invariance forbids any couplings 
of the triplet fields ($\chi$) to quarks.
For hypercharge $Y=2$ triplets, the neutral Higgs field $\chi^0$ 
can couple to leptons ($\nu\overline \nu$) via the following
Yukawa type interaction \cite{Schechter:1980gr}:
\footnote{Note that there is no such interaction
for $Y=0$ triplets, which are rendered fermiophobic as a consequence of 
gauge invariance}
\begin{equation}
h_{ij}\psi_{iL}^TCi\tau_2\Delta\psi_{jL}+h.c  \; .
\label{neutyuk}
\end{equation}
Here $h_{ij} (i,j=1,2,3)$ is an arbitrary coupling, $C$ is the Dirac charge
conjugation operator, $\psi_{iL}=(\nu_i, l_i)_L^T$ is a left-handed lepton
doublet, and $\Delta$ is a $2\times 2$ representation of the $Y=2$ complex
triplet fields:
\begin{equation}
\Delta
=\bma{cc}
\chi^+/\sqrt{2}  & \chi^{++} \\
\chi^0       & -\chi^+/\sqrt{2}
\ema  \; .
\end{equation}

The interaction described in eq.~(\ref{neutyuk}) has the virtue of
being able to provide neutrino masses and mixings consistent with
current neutrino oscillation data, {\sl without} invoking a
right-handed neutrino.  If the real part of the neutral triplet field
$\chi^{0r}$ acquires a vacuum expectation value (vev)
$\langle\chi^{0r}\rangle=b$, the following Majorana mass matrix
($m_{ij}$) for neutrinos is generated:
\begin{equation}
m_{ij}=\sqrt 2h_{ij}b \; .
\label{neutmass}
\end{equation}
Neutrino oscillation data constrain the product $h_{ij}b$, while
$h_{ij}$ is constrained directly by lepton flavour violating processes 
involving $\mu$ and $\tau$ e.g. $\mu\to e\gamma,\mu\to eee$ 
\cite{Cuypers:1996ia}. 
Hence it is clear that  $\chi^{0}$ is partially fermiophobic, with
a small coupling to neutrinos. 

We will consider the Higgs Triplet Model (HTM) of reference
\cite{Georgi:1985nv} in which a complex $Y=2$ triplet ($\Delta$) and a
real $Y=0$ triplet ($\xi^+,\xi^0,\xi^-$) are added to the SM
Lagrangian.  The HTM preserves $\rho=1$ at tree-level if the vev's of
both the neutral members are equal
$\langle\chi^0\rangle=\langle\xi^0\rangle=b$.  Taking the vev of the
Higgs doublet $\langle\Phi^0\rangle=a$, one has the following
expression for $m_W$ 
\begin{equation}
m_W^2={1\over 4}g^2(a^2+8b^2)\equiv {1\over 4}g^2 v^2
\end{equation}
where $v^2= 246^2$ ${\rm GeV}^2$.  It is convenient to define a
doublet-triplet mixing angle analogous to $\tan\beta$ in the 2HDM
\begin{equation}
\sin\theta_H= \left [ {8b^2\over a^2+8b^2} \right ]^{1/2} \; .
\end{equation}

In the HTM the physical Higgs boson mass spectrum is as follows
(in the notation of \cite{Gunion:1989ci})
\begin{equation}
   H_5^{\pm\pm},H_5^{\pm},H^0_5,H_3^\pm,H^0_3,H^0_1,H^{0'}_1 \; .
\end{equation}
The first five scalars are mass eigenstates, while the latter two can
mix in general; see below.  $H^0_1$ plays the role of the SM Higgs
boson and is composed of the real part of the neutral doublet field.
The eigenstate $H^{0'}_1$ is entirely composed of triplet fields and
is given by
\begin{equation}
  H^{0'}_1={1\over \sqrt 3}(\sqrt 2\chi^{0r}+\xi^0) \; .
\label{ferm}
\end{equation}

From the theoretical point of view, the size of the triplet vev $b$ is 
only constrained by the requirement that the doublet vev $a$ is 
sufficiently large to allow a perturbative top quark Yukawa coupling.
However, experimental constraints on $\sin\theta_H$ can be 
obtained by considering the effect of $H_3^\pm$ on processes such as
 $b\to s\gamma$, $Z\to b\overline b$ and $B-\overline B$ mixing
\cite{Kundu:1995qb}. 
Since $H^\pm_3$ has identical fermionic couplings to that of
$H^\pm$ in the 2HDM (Model I) with the replacement $\cot\beta\to 
\tan\theta_H$, one can derive the bound  $\sin\theta_H\le 0.4$.

In eq.~(\ref{ferm}) the $\chi^{0r}$ component in $H^{0'}_1$ couples to
$\nu\overline \nu$ via the $h_{ij}$ coupling. One can see from
eq.~(\ref{neutmass}) that the decay $H^{0'}_1\to \gamma\gamma$,
mediated by $W$ loops proportional to $b$, will dominate over
$H^{0'}_1\to \nu\overline \nu$ if $b$ is of the order of a few GeV.
Thus $H^{0'}_1$ is a candidate for a $h_f$, with BR$(H^{0'}_1\to
\gamma\gamma$) essentially equal to that of the benchmark $h_f$ model.
However, in general $H^0_1$ and $H^{0'}_1$ mix through the following
mass matrix written in the ($H^0_1$,$H^{0'}_1$) basis \cite{Gunion:1989ci}
\begin{equation}
{\cal M}
=\bma{cc}
8c^2_H(\lambda_1+\lambda_3)  & 2\sqrt 6s_Hc_H\lambda_3 \\
 2\sqrt 6s_Hc_H\lambda_3     & 3s_H^2(\lambda_2+\lambda_3)
\ema  v^2\; .
\end{equation}
Here $\lambda_i$ are dimensionless quartic couplings in the Higgs
potential and $s_H=\sin\theta_H$,$c_H=\cos\theta_H$.
The assumption that the $\lambda_i$ couplings are roughly the same order of
magnitude together with the imposition of the bound $s_H< 0.4$
results in very small mixing \cite{Akeroyd:1995ci}.
Moreover, $H^{0'}_1$ would be the lightest Higgs boson
in the HTM limit of small $s_H$, as stressed in \cite{Bamert:1993ah}.
In this paper we will study the
production process $qq'\to H_3^\pm H^{0'}_1$, assuming that
$H^{0'}_1$ is a fermiophobic Higgs with BRs equivalent to the benchmark 
$h_f$ model.

\subsection{Fermiophobic Higgs boson branching ratios}

\begin{figure}[h]
\begin{center} 
\includegraphics[width=10cm,angle=0]{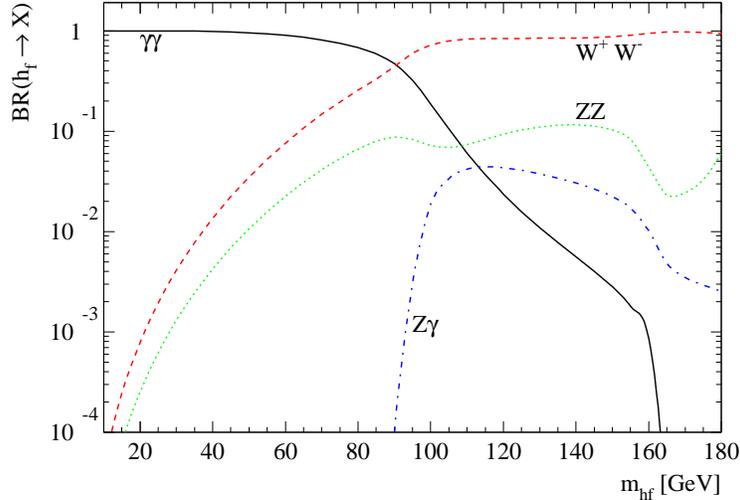}
\end{center}
\caption{Branching ratios of the largest decay modes of a fermiophobic
  Higgs boson assuming exact fermiophobia at tree-level. The branching
  ratio into $\gamma\gamma$ equals the $W^* W^*$ mode for
  $m_{h_f}\approx 90$ GeV and drops to $20$\% for $m_{h_f}=100$ GeV.}
\label{brhf}
\end{figure}

For the sake of illustration, we depict in Fig.\ \ref{brhf} the branching
ratios of a fermiophobic Higgs boson $h_f$ into $VV$ where $V$ can be 
either a $W$, $Z$ or $\gamma$.  In this figure we assumed that the $h_f$ 
couplings to fermions are absent and that $h_f\to \gamma\gamma$ 
is mediated solely by a $W$ boson loop,
\begin{center}
\vspace{-85pt} \hfill \\
\begin{picture}(300,140)(0,23) 
\DashLine(20,30)(60,30){3}
\Photon(60,30)(90,60){3}{6.5}
\Photon(90,0)(60,30){3}{6.5}
\Photon(90,60)(90,0){3}{8}
\Photon(90,60)(130,60){3}{6.5}
\Photon(90,0)(130,0){3}{6.5}
\Text(20,40)[]{$h_f$}
\Text(140,60)[]{$\gamma$}
\Text(140,0)[]{$\gamma$}
\Text(70,55)[]{$W^{\pm}$}
\DashLine(170,30)(210,30){3}
\PhotonArc(230,30)(20,0,360){3}{18}
\Photon(250,30)(280,60){3}{6.5}
\Photon(250,30)(280,0){3}{6.5}
\Text(170,40)[]{$h_f$}
\Text(290,60)[]{$\gamma$}
\Text(290,0)[]{$\gamma$}
\Text(230,60)[]{$W^{\pm}$}
\end{picture}  
\vspace{30pt} \hfill \\
\end{center}
giving rise to the following $h_f$ branching ratio into two photons,
\begin{equation}
\Gamma(h_f \to \gamma\gamma)=
{\alpha^2g^2\over 1024\pi^3}{m^3_{h_f}\over m_W^2} 
|F_1\cos\beta|^2 
\end{equation}
with $F_1=F_1(\tau)$, $\tau=4m^2_W/m^2_{h_f}$, a function given in 
\cite{Gunion:1989we}. We remind the reader that the $h_fWW$ coupling
normalized to the SM $\phi_0WW$ coupling satisfies
$\sin(\beta-\alpha)\rightarrow\cos\beta$ in the fermiophobic limit.

This gives rise to benchmark BRs which are used in the ongoing
searches to derive mass limits on $m_{h_f}$.  In practice, $h_f\to
\gamma\gamma$ can also be mediated by charged scalar loops: $H^\pm$ in
the 2HDM \cite{Barroso:1999bf},\cite{Brucher:1999tx} and
$H_3^\pm,H_5^\pm,H^{\pm\pm}_5$ in the HTM \cite{Gunion:1990dt}.
Although such contributions are suppressed relative to the $W$ loops
by a phase space factor, they can be important if the mixing angle
suppression for the $h_fWW$ coupling ($\cos\beta$) 
is quite severe {\em i.e. } the
scenario of interest in this paper. In our numerical analysis we will
assume the benchmark BRs given in Fig.\ \ref{brhf}.  One can see from
the figure that the loop induced decay mode $h_f \to \gamma\gamma$ is
dominant for $m_{h_f} \lsim 95$ GeV and drops below $0.1 \%$ for $h_f$
masses above 150 GeV. On the other hand, the decay channel $h_f \to
W^* W^*$ dominates for $m_{h_f} \gsim 95$ GeV, being close to 100\%
until the threshold for $h_f$ decay into two real $Z$'s is reached.

\subsection{The decay $H^\pm\rightarrow h_fW^*$}

\begin{figure}[h]
\begin{center} 
\includegraphics[width=15cm,angle=0]{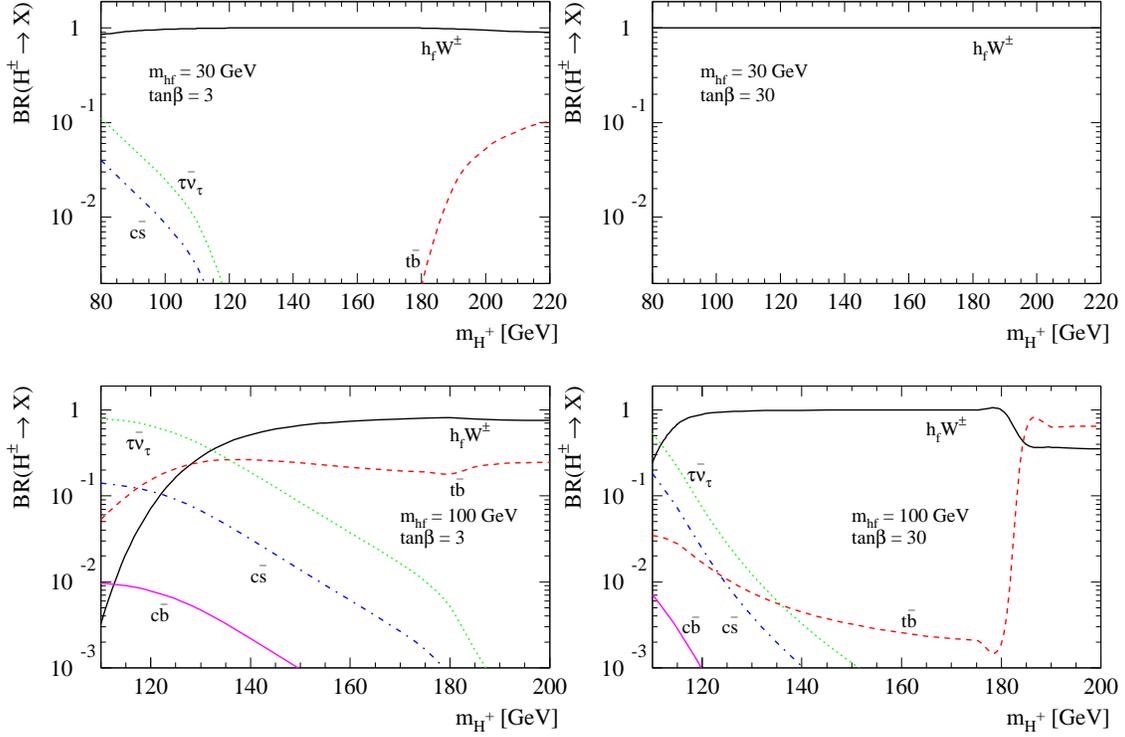}
\end{center}
\caption{The charged Higgs boson branching ratios into fermions
  $\tau\nu_\tau$ (green/dotted), $tb$ (red/dashed), $cs$
  (blue/dot-dashed), and $cb$ (magenta/solid), and $W^* h_f$ (black/solid)
  as a function of the charged Higgs boson mass for two different
  $\tan\beta$ and fermiophobic Higgs mass values.} 
\label{brch}
\end{figure}

The experimental signature of the process $qq'\to H^\pm h_f$ depends
on the decay modes of $H^\pm$. If $H^\pm$ decays to two fermions then
the signal would be of the type $\gamma\gamma + X$, which is
essentially the same as that assumed in the inclusive searches.
However, crucial to our analysis is the fact that the decay
$H^\pm\rightarrow h_fW^*$ may have a very large BR
\cite{Akeroyd:1998dt} in the 2HDM (Model~I). This is because the decay
width to the fermions ($H^\pm\to f'\overline f$) scales as
$1/\tan^2\beta$.  Similar behaviour occurs in the HTM
\cite{Akeroyd:1998zr} for the decay $H_3^\pm\to H^{0'}_1W^*$ with the
replacement $1/\tan^2\beta \to \tan^2\theta_H$.  Thus in the region of
$\tan\beta>10$ (or small $\sin\theta_H$) the fermionic decays of
$H^\pm$ are depleted.  This enables the decay $H^\pm\rightarrow
h_fW^*$ ($H^\pm_3\rightarrow H^{0'}_1W^*$) to become the dominant
channel {\sl even if} the mass difference $m_{H^\pm}-m_{h_f}$ is much
less than $m_W$. In Fig.~\ref{brch} we show the branching ratios of
the charged Higgs boson into fermions and $h_fW^*$ as a function of
$M_{H^\pm}$ for several values of $\tan\beta$ and $m_{h_f}$.  From the
right panels we see that in the large $\tan\beta$ regime the fermionic
decays are indeed suppressed.  Moreover, we also see that for light
fermiophobic Higgs bosons, where a $W$ boson can be produced on its
mass shell, the decay $H^\pm\rightarrow W^\pm h_f$ is essentially
100\% for any $\tan\beta$. On the other hand, for heavier fermiophobic
Higgs bosons, the fermionic decays can be the preferred decay channels
mainly for small $\tan\beta$.

%

\section{Phenomenology of $h_f$ at hadron colliders}

\subsection{$h_f$ production via the $VVh_f$ coupling}

Current searches at the Tevatron assume that production of $h_f$
proceeds via the $VVh_f$ coupling ($V=W,Z$) that originates from the
kinetic part of the Lagrangian. Run I searches utilized the process
$qq'\to Vh_f$ giving a signature of a $\gamma\gamma$ and a vector
boson \cite{Abbott:1998vv},\cite{Affolder:2001hx}.  The preliminary
Run II search is for inclusive $\gamma\gamma$ \cite{Melnitchouk:2005xv}
 and
is therefore sensitive to both $qq'\to Vh_f$ and the subdominant
vector boson fusion $qq'\to h_fqq$; see \cite{Mrenna:2000qh}.  Note
that $gg\to h_f$ via a fermion loop does not contribute to $h_f$
production.

In the 2HDM (Model I) the strength of the $VVh_f$ coupling
relative to the SM coupling $VV\phi^0$ is given by
\begin{equation}
    VVh_f\propto  \frac{1}{\sqrt{1+\tan^2\beta}}   \; .
\end{equation}
Hence the production mechanism $qq\to V^*\to Vh_f$ can be rendered 
completely ineffective for $\tan\beta>10$.  In the HTM the fermiophobic 
Higgs boson has a coupling size relative to $VV\phi^0$ given by
\begin{equation}
VVH^{0'}_1\propto \frac{2\sqrt 2}{\sqrt 3}~s_H \; .
\end{equation}
In direct analogy to the large $\tan\beta$ case of the 2HDM (Model I), 
a small $s_H$ would suppress the coupling $VVH^{0'}_1$ and consequently
deplete the $h_f V$ production. Hence it is of concern to consider
other production mechanisms which are unsuppressed in the above scenario.

\subsection{Associated $h_f$ production with $H^\pm$ and the
  multi-photon signature}

The production mechanism $qq'\to H^\pm h_f$ is complementary to that
of $qq'\to Vh_f$. This can be seen immediately from the explicit
expressions for the couplings. In the 2HDM (Model I) one has
\begin{equation}
    VH^\pm h_f \propto  \frac{\tan\beta}{\sqrt{1+\tan^2\beta} } \; ,
\end{equation}
while in the HTM
\begin{equation}
VH^\pm H^{0'}_1 \propto  \frac{2\sqrt 2}{\sqrt 3}~c_H \; .
\end{equation} 
Hence the above couplings are unsuppressed in the region of the parameter
space where the standard production mechanism $ q q' \to V h_f$
becomes ineffective. 
The larger coefficient for the $V H^\pm_3 H^{0'}_1$ coupling
is a consequence of the quantum number ($I,Y$) assignments in the 
HTM. 

To date complementary mechanisms have not been considered in
the direct fermiophobic Higgs searches at the Tevatron. As emphasized
in \cite{Akeroyd:2003bt}, \cite{Akeroyd:2003xi} a more complete search
strategy for $h_f$ at hadron colliders must include such production
processes in order to probe the scenario of fermiophobic Higgs bosons with
a suppressed coupling $h_fVV$. In the HTM one expects
$H^{0'}_1$ to be the lightest Higgs boson for small $\sin\theta_H$,
which further motivates a search in the complementary channel
$qq'\to H^\pm_3 H^{0'}_1$.

The experimental signature arising from the complementary mechanism $qq'\to
H^\pm h_f$ depends on the $H^\pm$ decay channel. In a large fraction of the
parameter space where the complementary mechanism $qq'\to H^\pm h_f$ is
important, the $H^\pm$ decay is dominated by $H^\pm \to h_f W^*$.
Consequently, this scenario would give rise to double $h_f$ production, 
with subsequent decay of $h_fh_f\to \gamma\gamma\gamma\gamma, 
VV\gamma\gamma$ and $VVVV$.  For light $h_f$ ($m_{h_f}\lsim 90 $ GeV), the 
signal $\gamma\gamma\gamma\gamma$ would dominate, as discussed in
\cite{Akeroyd:1998dt} at LEP, in \cite{Akeroyd:2003bt} for the Tevatron 
Run II and \cite{Akeroyd:2003xi} at the LHC and a Linear Collider. More 
specifically, the multi-photon signature arises in the portion of the 
parameters space where $m_{h_f}\lsim 90$ GeV, $m_{H^\pm}\lsim 200 $ GeV, 
and $\tan\beta > 1$ in the 2HDM Model I framework. In that region, 
$BR(h_f\rightarrow \gamma\gamma) \sim 1$ and $BR(H^\pm\rightarrow h_f 
W^{*\pm}) \sim 1$ as well, leading to a $4\gamma + \hbox{leptons or jets}$ 
signature.

As explained in \cite{Akeroyd:2003bt},
processes other than $qq'\to H^\pm h_f$
could give rise to a $4\gamma+X$ signal. 
One such mechanism is $q \bar{q} \to
A^0 h_f$, where $A^0$ is the heavy neutral pseudoscalar decaying
$A^0\to h_fZ^*$.
However, LEP already searched for $e^+e^-\to h_fA^0$
and set the bound $m_{h_f}+m_A>160$ GeV \cite{Abbiendi:2002yc}.
Thus any contribution from 
$qq\to A^0h_f$ will be phase space suppressed relative to that 
originating from $qq'\to H^\pm h_f$.
A similar argument applies to 
the production of a pair of charged Higgs bosons and its
subsequent decay into $h_f V^*$ pairs, {\em i.e.}  $q \bar{q} \rightarrow
Z^*, \gamma^* \rightarrow H^+ H^- \rightarrow h_f h_f W^+ W^-$ which
is phase space suppressed at Tevatron energies (2$m_{H^\pm}>180$ GeV
from direct $H^\pm$ searches). In the
minimal supersymmetric model the total rates for $H^+ H^-$ production are
enhanced in the large $\tan\beta$ regime through the Yukawa couplings of 
Higgs bosons to bottom quarks~\cite{hphm}, however in the 
2HDM Model~I and HTM, these 
Yukawa couplings are suppressed. The LHC would probably have much better 
prospects in these additional channels if all the above pair production
mechanisms were combined with the $H^\pm h_f$ associated
production in a fully inclusive multi-photon search. 
Since our analysis for the Tevatron we will
focus on $qq'\to H^\pm h_f$, which provides the best search potential
for the very light $h_f$ region because the phase space constraint
($m_{H^\pm}+m_{h_f}>100$ GeV) is  the least restrictive of the Higgs
pair production mechanisms.

\section{Multi-photon signal analyses}

We now present our analysis for the inclusive production of 
multi-photon final states which
may or may not be accompanied by extra leptons and/or jets, {\em i.e.}
the reaction
\[
  p\bar{p}\rightarrow h_f H^\pm\rightarrow h_f h_f W^\pm \rightarrow 
                              \gamma\gamma\gamma\gamma +X
\] 
at the Tevatron Run II. We focus our attention on two inclusive
final states; i) at least three photons ($>3\gamma$) and 
ii)  four photons ($4\gamma$). 
Only the ``1-prong'' tau lepton decays were considered.

In our analysis we evaluated the signal and standard model backgrounds
at the parton level. We calculated the full matrix elements using the
helicity formalism with the help of Madevent~\cite{mad}. We employed
CTEQ6L1 parton distributions functions~\cite{cteq} evaluated at the
factorization scale $Q_F=\sqrt{\hat{s}}$, where $\sqrt{\hat{s}}$ is
the partonic center-of-mass energy. Although QCD corrections increase
the tree--level cross section by a factor of around 1.3
\cite{Dawson:1998py}, we shall present results using the
tree--level cross sections only. Moreover we included momenta smearing
effects as given in \cite{Abbott:1998vv},
\cite{exper} and a detection efficiency of $85$\% per photon.

\begin{figure}[t]
\begin{center} 
\includegraphics[width=17cm,angle=0]{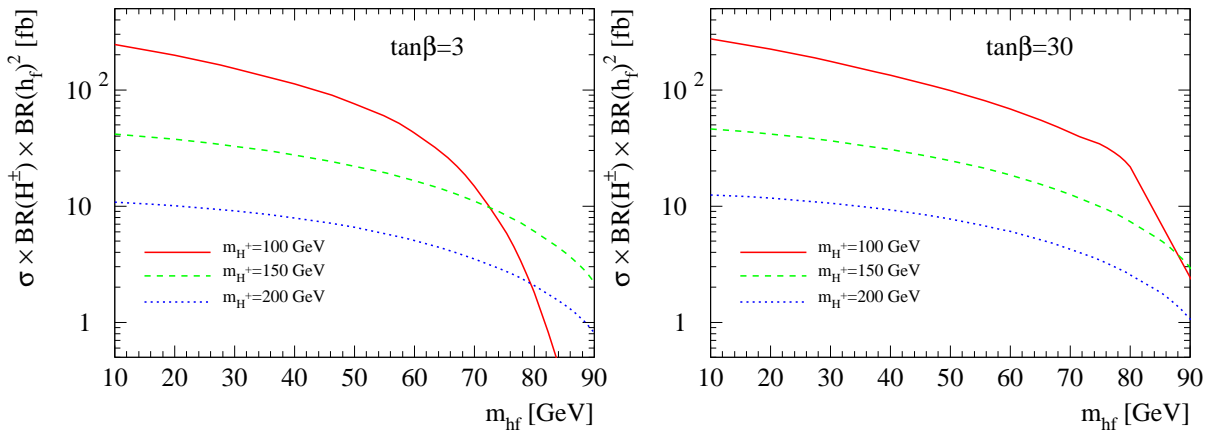}
\end{center}
\caption{Total production cross sections times branching ratios of
  $h_f\rightarrow \gamma\gamma$ and $H^\pm\rightarrow W^\pm h_f$
  for $p\bar{p}\rightarrow h_f H^\pm\rightarrow h_f h_f +
  W^\pm\rightarrow \gamma\gamma\gamma\gamma+W^\pm$ before cuts 
  at the Tevatron Run II in femtobarns. The values of $\tan\beta$ and
  $m_{H^\pm}$ are as indicated in the figure. }
\label{sigbr}
\end{figure}

We present in Fig.\ \ref{sigbr} the total signal cross section times
the branching ratios of $H^\pm \to h_f W^\pm$ and $h_f \to \gamma
\gamma$ for the complementary process $q q'\to H^\pm h_f$; these
results were obtained without cuts and detection efficiencies. In the
left (right) panel we present the $4\gamma$ production cross section
before cuts as a function of $m_{h_f}$ for three different values of
$m_{H^\pm}$ and $\tan \beta = 3$ (30). In the left panel, where
$\tan\beta=3$, the upper curve ($m_{H^\pm} = 100$~GeV) shows the strongest
effect of the phase space suppression of the decay $H^\pm \to W^* h_f$
for $m_{h_f} \gsim 60$ GeV. This cross section reduction can be
partially compensated by the increase in $\tan\beta$ as shown in the
right panel. From the figure it is evident that this process
will produce a large number of events before cuts over a large
fraction of the parameter space.

Potential SM backgrounds for the multi-photon signature of fermiophobic
Higgs bosons are:
i) the three and four photon production $ p \bar{p} \to
\gamma \gamma \gamma (\gamma)$, ii) three photons and a $W$ production $p
\bar{p} \to \gamma \gamma \gamma W$, and iii) the associated production of
two or three photons and a jet where the latter is misidentified as
a photon $ p \bar{p} \to \gamma \gamma (\gamma) j (\to \gamma + X)$.
We verified that after cuts and taking into account a $P(j \to
\gamma)=4\times 10^{-4}$~\cite{exper} photon misidentification probability
the total SM background amounts to $3.8$ events for an integrated
luminosity of 2 fb$^{-1}$.  Therefore the complementary process for
the fermiophobic Higgs search has the great advantage of being
extremely clean for a large portion of the 2HDM and HTM parameter space.
In contrast, the ongoing search for inclusive $\gamma\gamma + X$
\cite{Melnitchouk:2005xv}
suffers from a sizeable background originating from QCD jets
faking photons. For the exclusive channel ($\gamma\gamma + V$) 
the background is considerably smaller but still not negligible
\cite{Abbott:1998vv},\cite{Affolder:2001hx}.

\subsection{Searches at the Tevatron Run II}

The multi-photon topology is privileged concerning the level of
background, which is small in the SM after mild cuts.
Consequently, we imposed a minimum set of cuts on the final state
particles, in order to guarantee their identification and 
isolation. Further studies could optimize the search strategy. 
We required the events to 
possess central photons with enough transverse energy to assure their 
proper identification
\begin{equation}
       E^\gamma_T >  15\; \hbox{GeV}\;\;\; , \;\;\;  |\eta^\gamma|  
       <  1.0\;\; ,
\label{eq:set1}
\end{equation}
and isolated from the other particles in the final state ($X=$ charged
lepton or jet) with a transverse energy in excess of 5 GeV
\begin{equation}
\Delta R_{\gamma\gamma} >  0.4\;\;\;\; ,\;\;\;\; \Delta R_{\gamma X} >
0.4 \; .
\label{eq:set1a}
\end{equation}
Notice that the high $p_T$ central photons is enough to guarantee the
trigger of these events \cite{exper}. These cuts are very effective
against the backgrounds from continuous $\gamma\gamma\gamma + X$
production which occur mainly through photon and gluon 
bremsstrahlung emission from initial and/or final state quarks, and gluon
splitting to collinear quarks. We have checked that $4\gamma + X$
topologies give a negligible contribution after imposing the cuts.

\begin{figure}[t]
\begin{center} 
\includegraphics[width=17cm,angle=0]{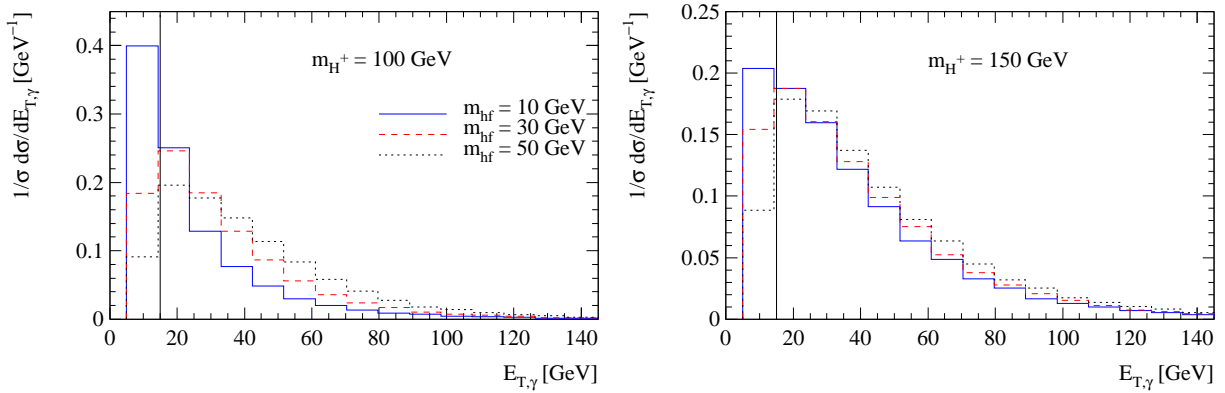}
\end{center}
\caption{Normalized transverse energy distributions (in GeV) of photons
  for two different charged Higgs boson masses and three different
  fermiophobic Higgs masses. The vertical solid lines indicate the
  $E^\gamma_T$ cut in eq.~(\ref{eq:set1}).}
\label{dist_etx}
\end{figure} 

In order to understand the effect of these cuts on the signal we studied
some kinematical distributions. We present in Fig.\ \ref{dist_etx} the
normalized transverse energy distribution of the final state photons
for several values of Higgs masses and $\tan\beta = 30$. As one can
see, the $E_T^\gamma$ spectrum peaks around $\lsim~ m_{h_f}/2$ and the
spectrum at low $E_T^\gamma$ decreases as the fermiophobic Higgs
becomes heavier. We can also learn that the $E^\gamma_T$ distribution
becomes harder as the charged Higgs mass increases. Thus, the
transverse energy cut in eq.~(\ref{eq:set1}) attenuates more the light
fermiophobic Higgs signal. 

\begin{figure}[t]
\begin{center} 
\includegraphics[width=17cm,angle=0]{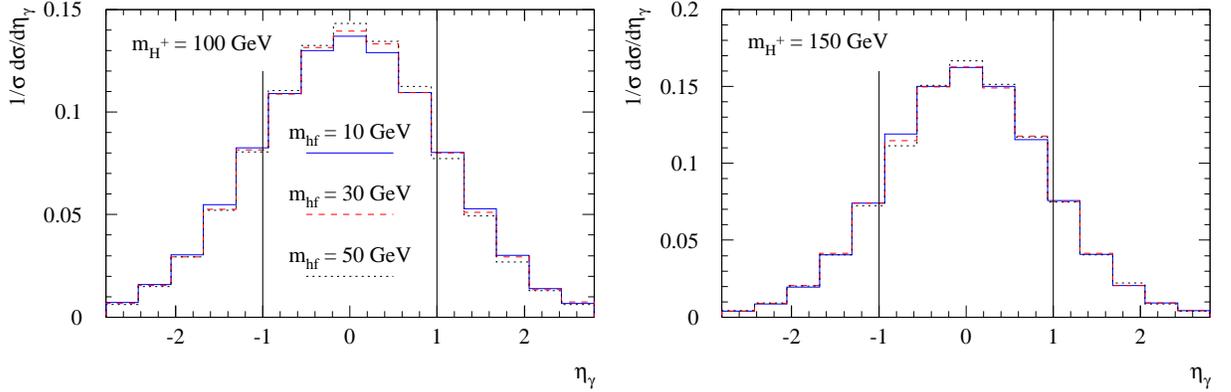}
\end{center}
\caption{Normalized rapidity distributions of photons for two
  different charged Higgs boson masses and three different
  fermiophobic Higgs masses. The vertical solid lines indicate the 
  $\eta_\gamma$ cut of eq.~(\ref{eq:set1}).}
\label{dist_yx}
\end{figure}

\begin{figure}[h]
\begin{center} 
\includegraphics[width=17cm,angle=0]{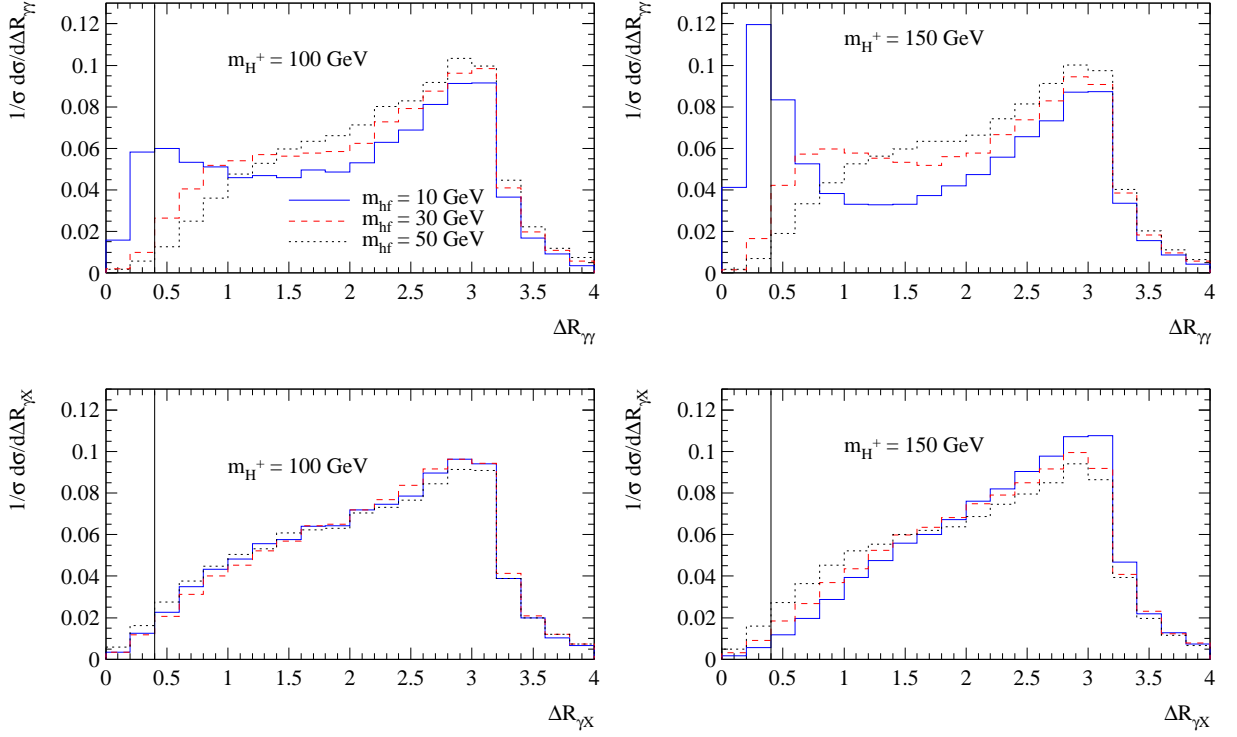}
\end{center}
\caption{Normalized cone variable distributions between two photons
  (upper panels), a photon and a charged particle or jet (lower
  panels) for the same parameter used in Fig.\ \ref{dist_etx}.  The
  vertical solid lines indicate the $\Delta R$ cuts 
  in eq.~(\ref{eq:set1a}).
}
\label{dist_rxx}
\end{figure}

\begin{figure}[h]
\begin{center} 
\includegraphics[width=15cm,angle=0]{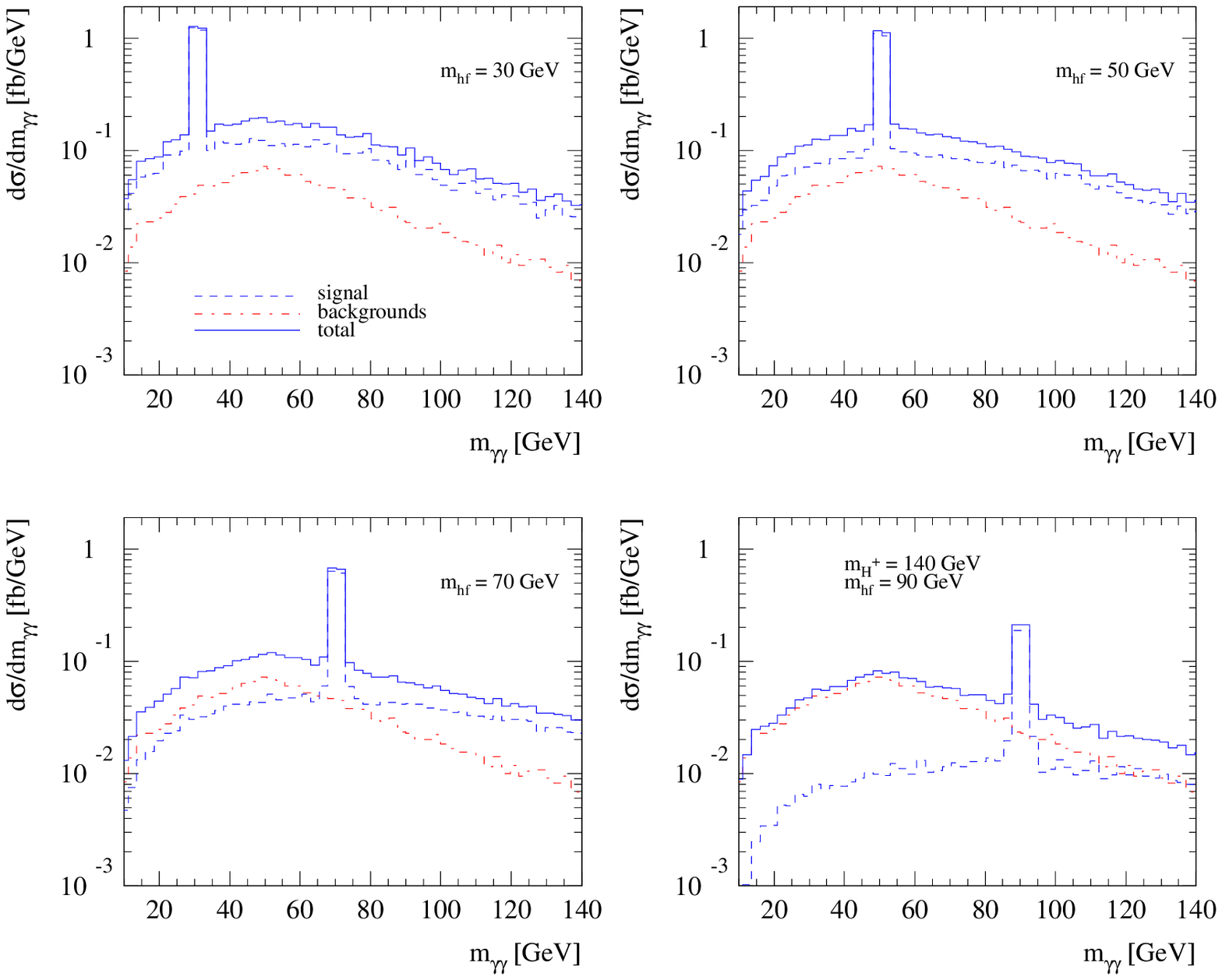}
\end{center}
\caption{The photon--photon invariant mass spectrum for 
  $m_{H^\pm}= 150$ GeV, $m_{h_f}=30$, $50$, $70$ GeV, 
  and $m_{H^\pm}= 140$ GeV, $m_{h_f}=90$ GeV at the upper left, upper
  right, bottom left, and bottom right panels, respectively. Also
  shown are the sum of all backgrounds consisting of $3\gamma + X$
  final states. In these 
  plot we entered the invariant mass of all photon pair possible 
  combinations. In all cases we set $\tan\beta=30$. }
\label{dist_mx}
\end{figure}

Figure \ref{dist_yx} contains the photon
rapidity distribution for the same parameters used in Fig.\
\ref{dist_etx}. The rapidity distribution of the photons stemming from
the fermiophobic Higgs decay peaks around zero. However, there is a
sizeable contribution from high rapidity photons. For heavier charged
Higgs bosons the rapidity distribution is more central.  The hardest
cut that we applied is the requirement that the absolute value 
of the photon rapidity be
smaller than unity, and its effect is
rather insensitive to the neutral Higgs mass.  On the other hand, 
the separation cuts in eq.~(\ref{eq:set1a}) have little effect on the 
signal cross section as shown in Fig.\ \ref{dist_rxx}, with perhaps 
the exception of very small fermiophobic Higgs masses. Notice that we 
did not introduce any cut on the photon--photon invariant masses. 
Certainly if a signal is observed the photon pair invariant mass will 
display a clear peak at $m_{h_f}$ even after adding all the possible 
photon pair combinations and backgrounds; see Fig.~\ref{dist_mx}.

\begin{figure}[h]
\begin{center} 
\includegraphics[width=10cm,angle=0]{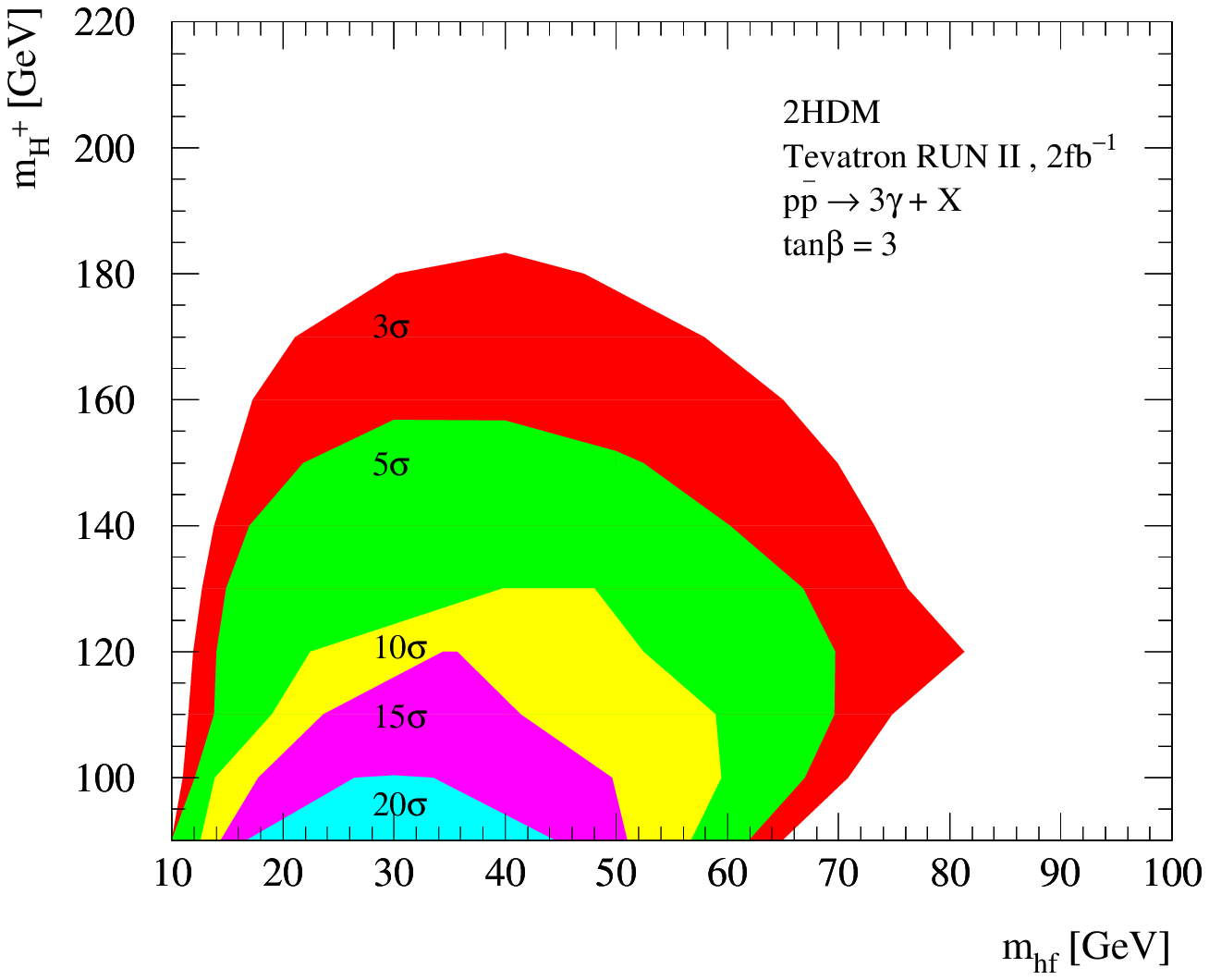} \\
\includegraphics[width=10cm,angle=0]{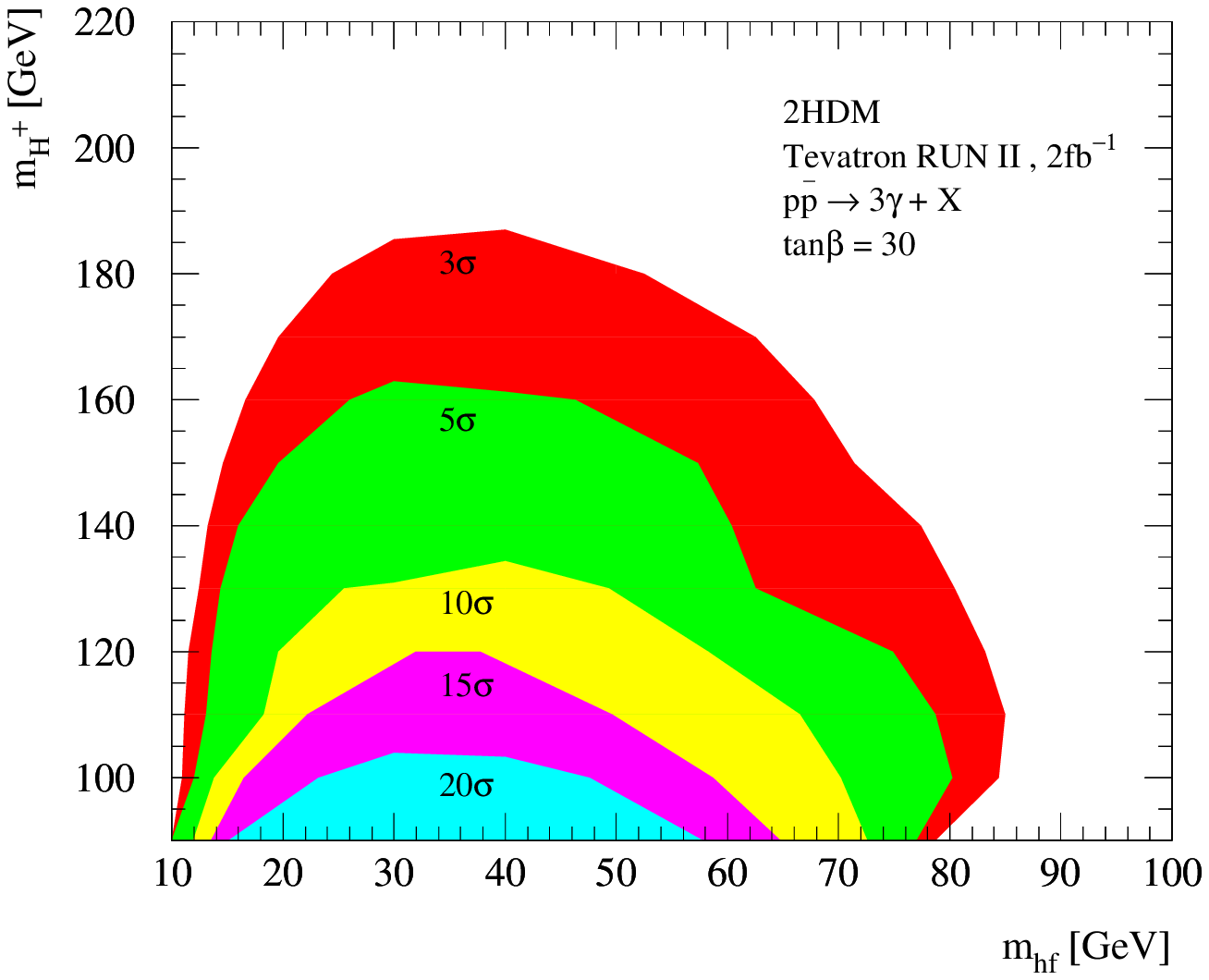}
\end{center}
\caption{ Expected signal statistical significance
  presenting three of more photons
  in the $m_{H^\pm} \otimes m_{h_f}$ plane assuming an
  integrated luminosity of 2 fb$^{-1}$ at the Tevatron RUN II. We
  assumed the 2HDM Model I and took $\tan\beta = 3$ (30) in the upper
  (lower) panel.  }
\label{fig:cont1}
\end{figure}

We display in Figure \ref{fig:cont1} 
the region in the plane $m_{H^\pm} \otimes m_{h_f}$ where at least a
3$\sigma$ signal can be observed,
exhibiting three or more photons, in the framework of the 2HDM Model I 
for an integrated luminosity of 2 fb$^{-1}$. Statistical significance
$\sigma$ is defined by $\sigma=S/\sqrt{B}$, 
where $S(B)$ is the number of signal (background)
events after applying cuts and efficiency factors. 
A few comments are in order. First of all, the expected number 
of events diminishes for small $h_f$ masses since fewer events pass the 
$E^\gamma_T$ cut in eq.~(\ref{eq:set1}) as can be seen from 
Fig.\ \ref{dist_etx}. Secondly, the shape of the region presenting at 
least 3$\sigma$ (5$\sigma$ or 10$\sigma$) significance in the large 
$h_f$ mass region is the result of 
a competition between the phase space suppression of the cross section as 
$m_{H^\pm}$ increases for fixed $m_{h_f}$ and the growth of the $H^\pm 
\to W^\pm h_f$ branching ratio; see Fig.  \ref{brch}. Furthermore, for a 
fixed number of events, the optimum reach in $m_{H^\pm}$ takes place for 
$m_{h_f} \simeq 30$--$40$ GeV.  This is a consequence of the combined 
effects of cuts and phase space suppression as we have already discussed.

As one can see from the upper panel
in Fig.\ \ref{fig:cont1}, even in the low
$\tan\beta$ region the reach of the Tevatron RUN II is quite impressive in
this scenario. If no events were observed above the backgrounds 
at RUN II, a large fraction of
the $m_{H^\pm} \otimes m_{h_f}$ plane would be excluded at the 3$\sigma$
level. The situation
improves slightly for larger $\tan\beta$ as can be seen from the lower panel of
Fig.\ \ref{fig:cont1}. Importantly, the expected number 
of events is rather large in the region $m_{h_f} \lsim 70 $ GeV and 
$m_{H^\pm} \lsim 150$ GeV, such that it will be possible to 
reconstruct the 
$h_f$ mass from the photon--photon invariant mass distribution; see 
Fig.\ \ref{dist_mx}.

\begin{figure}[h]
\begin{center} 
\includegraphics[width=10cm,angle=0]{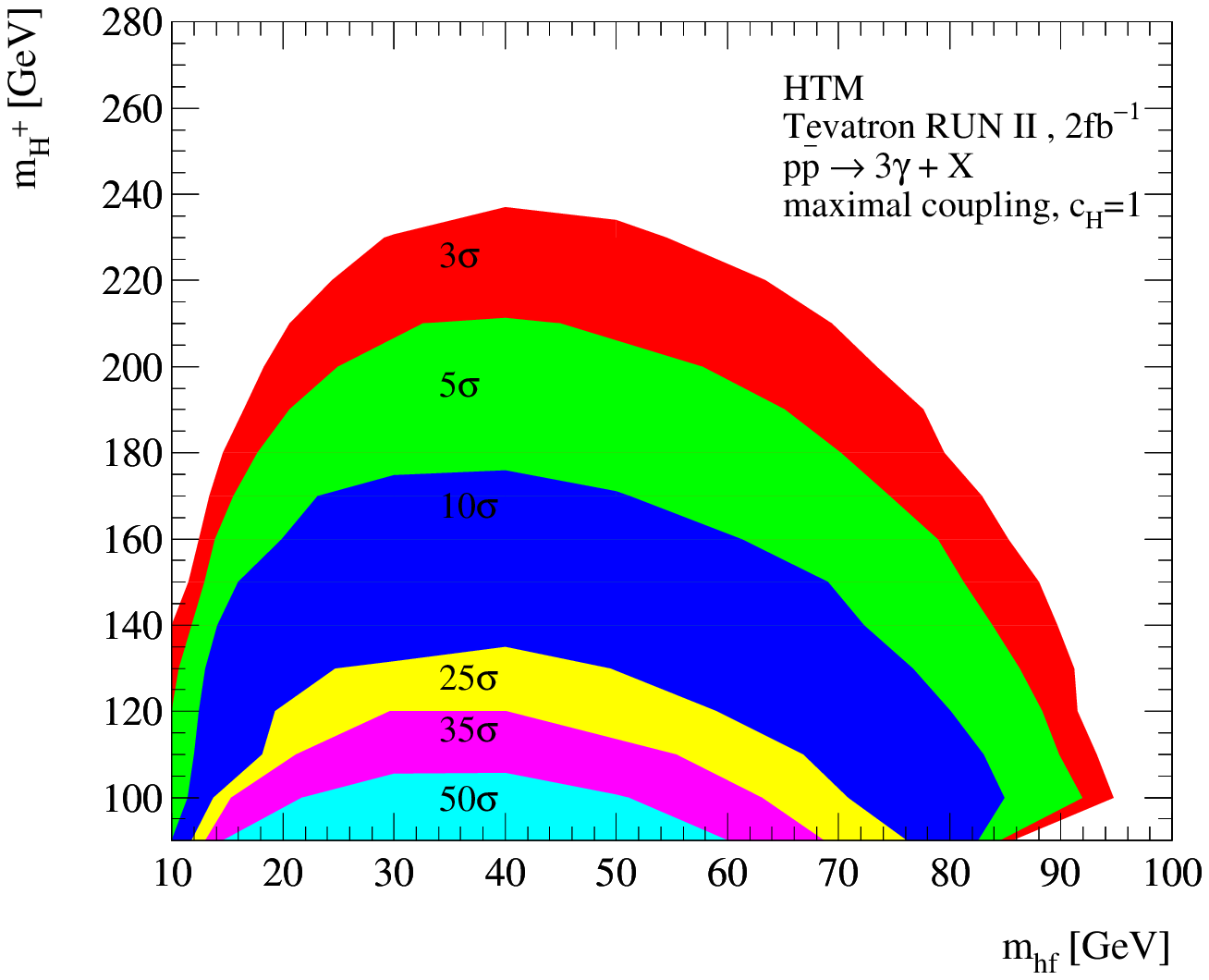} \\
\includegraphics[width=10cm,angle=0]{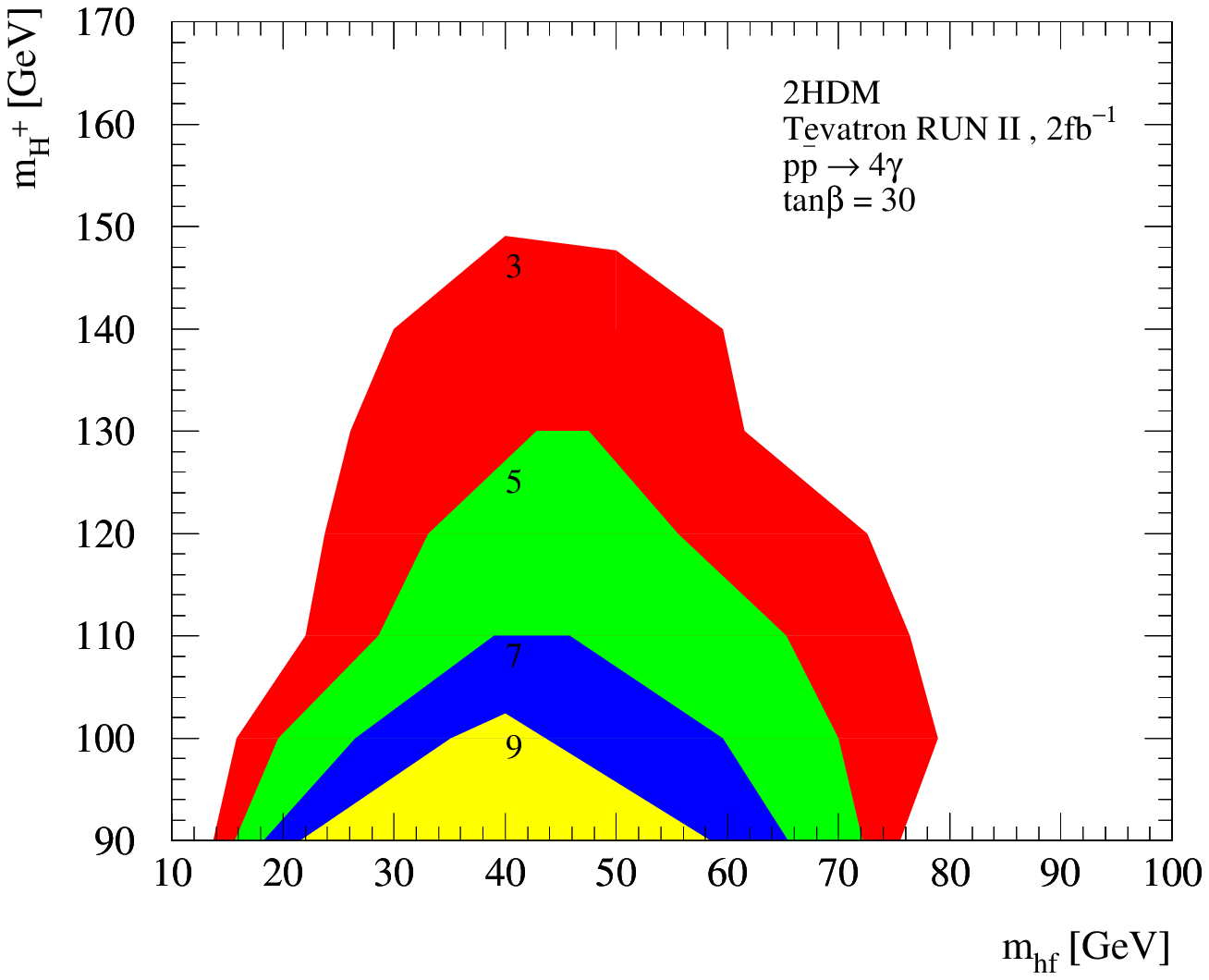}
\end{center}
\caption{In the upper panel we display the expected signal statistical
  significance
  containing three of more photons in the $m_{H^\pm} \otimes m_{h_f}$ plane 
  in the HTM framework, assuming an integrated luminosity of 2 fb$^{-1}$ 
  at the Tevatron RUN II. In the lower panel we present the expected number 
  of events presenting four photons in the $m_{H^\pm} \otimes m_{h_f}$ 
  plane for the 2HDM Model I and assuming $\tan\beta = 30$. }
\label{fig:cont2}
\end{figure}

For comparison, we present in Fig.~\ref{fig:cont2} the expected signal
significance of events containing three or more photons after cuts 
for the HTM.  
In our numerical analysis we take $c_H=1$ as a benchmark value,
and the signal significance for other values of $c_H$ can be obtained
by simply rescaling the displayed numbers. 
From the bound $s_H < 0.4$ one obtains $c_H > 0.9$.
In the exact $c_H=1$ limit
(i.e. triplet vev $b=0$) the neutrinos would not receive a mass
at tree-level (see eq.~\ref{neutmass}).
Extremely small $s_H < 10^{-9}$  would require 
non-perturbative values of
$h_{ij}$ to generate realistic neutrino masses.
We are interested in the interval $0.9 < c_H < 0.99$ (corresponding
to GeV scale triplet vev) in which $H^{0'}_1$ decays
primarily to photons in the detector, and
neutrino mass is generated with a very small $h_{ij}\sim 10^{-10}$.

It is clear that a larger region 
of the $m_{h_f} \otimes m_{H^\pm}$ parameter space  
can be probed in the HTM than in the 2HDM. In 
fact, at the 3$\sigma$ level RUN II will be able to exclude Higgs masses up to 
$ m_{H^\pm} \lsim 240 $ GeV or $m_{h_f} \lsim 100$ GeV. In order to 
understand the signal suppression if one requires an inclusive state 
containing four photons to pass our cuts, we present in 
Fig.~\ref{fig:cont2} lower panel the expected number of events for the 
2HDM, assuming $\tan\beta=30$ and an integrated luminosity of 2 
fb$^{-1}$. As expected, not only the reach in $m_{H^\pm}$ gets reduced to
$m_{H^\pm}\lsim 150$ GeV at 95\% C.L., but also the low and high $h_f$ 
mass regions become substantially depleted.

\section{Conclusions}

Higgs bosons with very suppressed couplings to fermions 
($h_f$) can arise in 
various extensions of the Standard Model (SM) such as 
the Two Higgs Doublet Model (2HDM) Type I or Higgs Triplet Model (HTM).  
Their conventional production mechanism at hadron colliders
$qq'\to W^\pm h_f$ can be severely suppressed 
by either large $\tan\beta$ or small triplet vacuum expectation value.
In this scenario the complementary channel $p \bar{p} \to H^\pm h_f$
is maximal and provides an alternative production mechanism. We 
studied the reaction $qq'\to H^\pm h_f$ followed by 
the potentially important decay $H^\pm\to h_fW^*$. We performed a Monte 
Carlo simulation of the detection prospects for a light $h_f$ where the 
branching ratio into photon pairs is dominant, which gives rise 
to multi-photon signatures with very low SM background. 
We showed that if a signal 
containing at least three photons is not seen at the Tevatron RUN II then 
a large portion of the $m_{H^\pm}$ \emph{versus} $m_{h_f}$ plane can be 
excluded both in the small and large $\tan\beta$ regimes of the 2HDM. 
Conversely, if a signal {\it were} observed then $>50$ events are 
expected for a light $H^\pm$ and $h_f$, which would 
allow further detailed phenomenological studies.

\section*{Acknowledgements}  

We would like to thank Oleksiy Atramentov for discussions concerning
the SM backgrounds.
This research was supported in part by Funda\c{c}\~{a}o de Amparo \`a
Pesquisa do Estado de S\~ao Paulo (FAPESP), by Conselho Nacional de
Desenvolvimento Cient\'{\i}fico e Tecnol\'ogico (CNPq), and by 
Conicyt grant No.~1040384.


\end{document}